\documentclass{aastex}
\usepackage{spr-astr-addons}
\usepackage{url}

\RequirePackage{color}

\begin{document}

\title{CCD BV and 2MASS photometric study of the open cluster NGC 1513}
\shorttitle{Open Cluster NGC 1513}
\shortauthors{S. Bilir, T. G\"uver, I. Khamitov, T. Ak, S. Ak, K. B. Co\c skuno\u glu, E. Paunzen, E. Yaz}

\author{S. Bilir \altaffilmark{1}}
\altaffiltext{1}{Istanbul University, Faculty of Science, Department 
of Astronomy and Space Sciences, 34119 University, Istanbul, Turkey\\
\email{sbilir@istanbul.edu.tr}}
\and
\author{T. G\"uver\altaffilmark{2}} 
\altaffiltext{2}{University of Arizona, Department of Astronomy, 933 N. Cherry 
Ave., Tucson, AZ 85721}
\and
\author{I. Khamitov\altaffilmark{3}} 
\altaffiltext{3}{T\"UB\.ITAK National Observatory, Akdeniz University Campus, 07058 
Antalya, Turkey}
\and
\author{T. Ak\altaffilmark{1,3}}
\altaffiltext{1}{Istanbul University, Faculty of Science, Department 
of Astronomy and Space Sciences, 34119 University, Istanbul, Turkey\\}
\altaffiltext{3}{T\"UB\.ITAK National Observatory, Akdeniz University 
Campus, 07058 Antalya, Turkey}
\and
\author{S. Ak\altaffilmark{1}}
\altaffiltext{1}{Istanbul University, Faculty of Science, Department 
of Astronomy and Space Sciences, 34119 University, Istanbul, Turkey\\}
\and
\author{K. B. Co\c skuno\u glu\altaffilmark{1}}
\altaffiltext{1}{Istanbul University, Faculty of Science, Department 
of Astronomy and Space Sciences, 34119 University, Istanbul, Turkey\\}
\and
\author{E. Paunzen\altaffilmark{1}}
\altaffiltext{4}{Institut f\"ur Astronomie der Universit\"at Wien, T\"urkenschanzstr. 
17, 1180 Wien, Austria}
\and
\author{E. Yaz\altaffilmark{1}}
\altaffiltext{1}{Istanbul University, Faculty of Science, Department 
of Astronomy and Space Sciences, 34119 University, Istanbul, Turkey\\}

\begin{abstract}
We present CCD BV and JHK$_{s}$ 2MASS photometric data for the open cluster NGC 1513. We observed 609 stars in the direction of the cluster up to a limiting magnitude of $V\sim19$ mag. The star count method shows that the centre of the cluster lies at $\alpha_{2000}=04^{h}09^{m}36^{s}$, $\delta_{2000}=49^{\circ}28^{'}43^{''}$ and its angular size 
is $r=10$ arcmin. The optical and near-infrared two-colour diagrams reveal the colour excesses in the direction of the cluster as $E(B-V)=0.68\pm0.06$, $E(J-H)=0.21\pm0.02$ and $E(J-K_{s})=0.33\pm0.04$ mag. These results are consistent with normal interstellar extinction values. Optical and near-infrared Zero Age Main-Sequences (ZAMS) provided an average distance modulus of $(m-M)_{0}=10.80\pm0.13$ mag, 
which can be translated into a distance of $1440\pm80$ pc. Finally, using Padova isochrones we determined the metallicity and age of the cluster as $Z=0.015\pm 0.004$ ($[M/H]=-0.10 \pm 0.10$ dex) and $\log (t/yr) = 8.40\pm0.04$, respectively.

\end{abstract}

\keywords{Galaxy: Open Cluster and associations: individual: NGC 1513 stars: interstellar extinction} 

\section{Introduction}
Systematic studies of open clusters help understand the galactic structure 
and star formation processes as well as stellar evolution. By utilizing 
colour-magnitude diagrams of the stars observed in the optical/near-infrared 
(NIR) bands, it is possible to determine the underlying properties of open 
clusters such as age, metal abundance and distance. In this study, we 
continue this effort by analyzing the optical and NIR data of the open 
cluster NGC 1513.

The young open cluster NGC 1513 = C 0406+493 ($\alpha_{2000}$ $=04^{h} 09^{m} 
10^{s}$, $\delta_{2000}=49^{\circ}31^{'} 00^{''}$, ${\it l}=152^{\circ}.50$, 
${\it b}=-1^{\circ}.66$) is classified as a Tr\"umpler class II 1m, a moderately 
rich cluster with a low central concentration. NGC 1513 was first studied 
astrometrically by \citet{B58a} who determined the proper motions 
using a single pair of plates with an epoch difference of 55 years. \citet{BD60}
 published photographic and photovisual magnitudes of 
49 stars from Bronnikova's (1958b) list. \citet{DH88} 
obtained the first photoelectric UBV magnitudes of 31 stars and photographic 
RGU magnitudes for 116 stars in the cluster region. They determined colour 
excess, distance and the age of the cluster from RGU data as $E(G-R)=0.93$ mag, 
1320 pc and $\log (t/yr)=8.18$, respectively. To test stellar evolution models, \citet{F98} 
studied NGC 1513 using BV photometry. \citet{F02} studied 
the astrometric data-sets of 333 stars in the direction of the cluster, and 
showed that 33 of those stars are most probably cluster members. Using 
BV photometric data, they also determined the cluster's metal abundance to be about 
solar metallicity and its age as $\log (t/yr)= 8.40$. \citet{MN07} studied 42 open 
clusters using CCD BV photometry and obtained the structural and astrophysical 
parameters for NGC 1513. According to their measurements the cluster's colour excess, distance 
modulus and age are $E(B-V)=0.76$ mag, $(m-M)=12.96$ mag and $\log (t/yr)=7.4$, 
respectively.

In this study we observed the open cluster NGC 1513 with BV filters, and matched our results 
with 2MASS photometry. We determined the colour excesses in optical and 
near-infrared region. Then, we used optical and near-infrared data to 
obtain the cluster's distance modulus, metal abundance and age.

The paper is organized as follows. We present the observations
and data reductions in Section 2, while in Section 3, we describe
the data analysis. Finally, Section 4 contains the conclusions of
our study.

\section{Observations and Data Reduction}
\subsection{Optical Data}
CCD BV photometric observations of NGC 1513 were made in 8th and 9th of October 2004 at the T\"{U}B\.{I}TAK National Observatory (TUG) using the 1.5-m Russian-Turkish Telescope RTT150 and ANDOR
DW436 CCD camera (back illuminated, 2k$\times$2k pixels, 13.5$\times$ 13.5 $\mu$m). The resulting image on the CCD has a field of view of  $8' \times 8'$. Since the spatial diameter of NGC 1513 is approximately $15' \times 15'$ we divided the field into four equal subfields and created a mosaic image from these. Coordinates of the subfields and observation summary are given in Table 1 and the subfields are shown in Fig. 1.

\begin{table*}
\setlength{\tabcolsep}{3pt} 
\center
\caption{Coordinates of the subfields and log of observations with dates, airmass and exposure times for each filter.  
Exposure times are given in seconds.}
\begin{tabular}{ccccccccc}
\hline
Sub-& &  & \multicolumn{3}{c}{B Band} & \multicolumn{3}{c}{V Band} \\
field &$\alpha_{2000}$  & $\delta_{2000}$& Exposure & Date & Airmass & Exposure & Date & Airmass \\
\hline
F1  & 04 10 10 & +49 33 20 & 30$\times$3 & 10/09/2004 & 1.036 & 10$\times$3 & 10/08/2004 & 1.257 \\
F1  & 04 10 10 & +49 33 20 &600$\times$3 & 10/09/2004 & 1.030 & 30$\times$3 & 10/08/2004 & 1.228 \\
F2  & 04 09 32 & +49 33 20 & 30$\times$3 & 10/09/2004 & 1.025 & 10$\times$3 & 10/08/2004 & 1.188 \\
F2  & 04 09 32 & +49 33 20 &600$\times$3 & 10/09/2004 & 1.026 & 30$\times$3 & 10/08/2004 & 1.168 \\
F3  & 04 10 10 & +49 27 15 & 30$\times$3 & 10/09/2004 & 1.089 & 10$\times$3 & 10/08/2004 & 1.091 \\
F3  & 04 10 10 & +49 27 15 &600$\times$3 & 10/09/2004 & 1.063 & 30$\times$3 & 10/08/2004 & 1.079 \\
F4  & 04 09 32 & +49 27 15 & 30$\times$3 & 10/09/2004 & 1.097 & 10$\times$3 & 10/08/2004 & 1.137 \\
F4  & 04 09 32 & +49 27 15 &600$\times$3 & 10/09/2004 & 1.037 & 30$\times$3 & 10/08/2004 & 1.121 \\
\hline
\end{tabular}
\end{table*}

\begin{figure*}
\centering
   \includegraphics[scale=0.75]{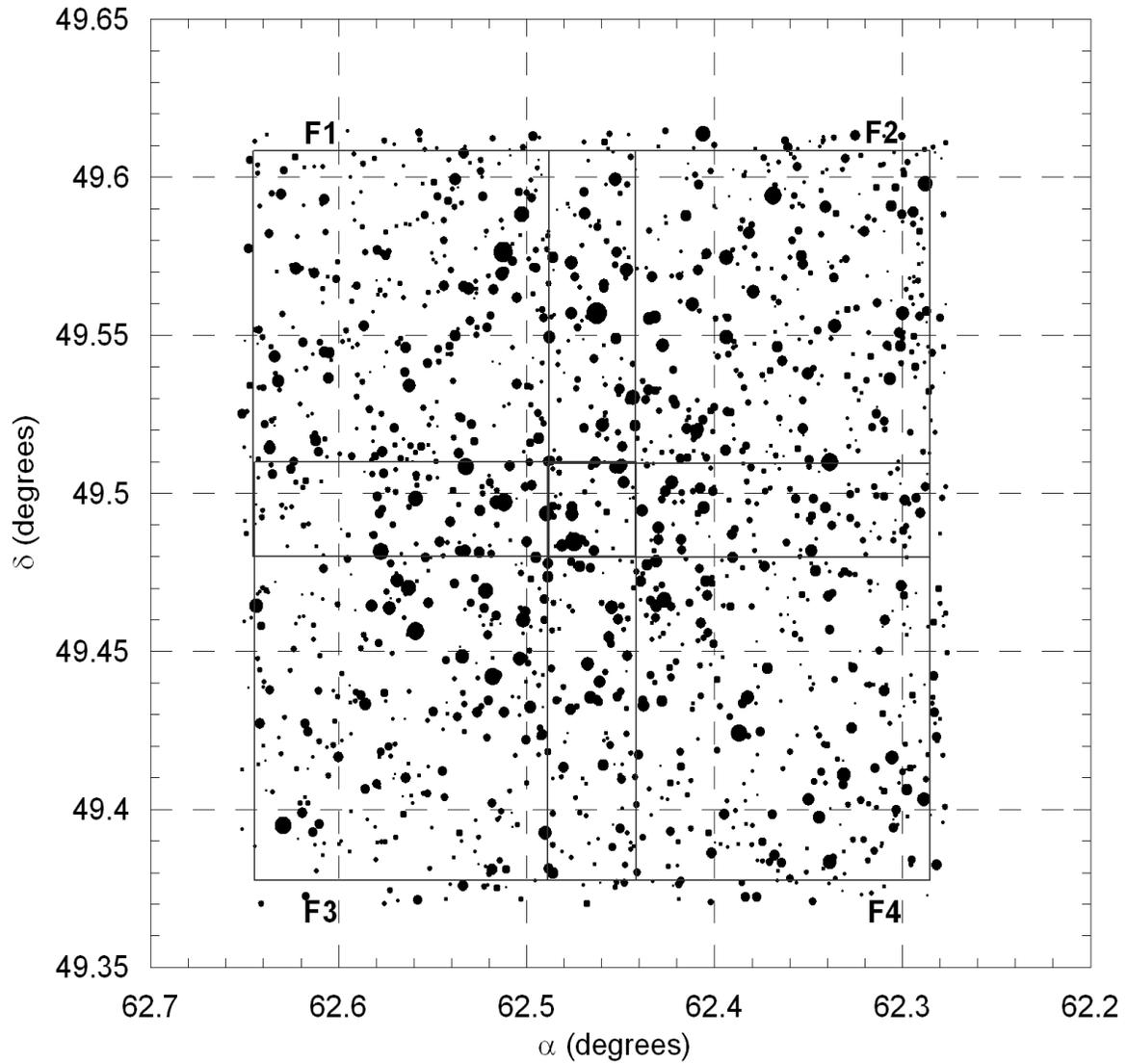}
   \caption{Finding chart of the stars in the field of NGC 1513. Four subfields (F1, F2, F3 and F4) were also shown. Filled circles of different sizes represent brightness of the stars. Smallest size denotes stars of $V \sim 19$ mag.}
\label{fields}
\end{figure*}

For each subfield we obtained six images in B and V bands, which are then combined for each filter and median images are used for further analysis. Obtained images were reduced using the computing
facilities available at TUG, Turkey. The standard IRAF\footnote{http://iraf.noao.edu} routines were utilized for prereduction, and the IRAF version of the DAOPHOT package \citep{S87,S92} was used with a quadratically varying point spread function (PSF) to derive positions and magnitudes for the stars. To determine the PSF, we used several well-isolated stars in the entire frame. Output catalogues for each frame were aligned in position and magnitude, and final (instrumental) magnitudes were computed as weighted averages of the individual values. Magnitudes of the stars brighter than $V=10$ could not be measured due to saturation of the detector pixels.

The instrumental b and v magnitudes were transformed into standard Johnson B and V magnitudes using fitting coefficients derived from observations of the standard field stars whose magnitudes were given by \citet{DH88} and \citet{F98} for photoelectric UBV and CCD UBV photometry, and taking airmass corrections into account. 15 and 16 standard stars given by \citet{DH88} and \citet{F98} have a magnitude and colour range of $12<V<18$ and $0.55<B-V<1.70$ mag. Errors of magnitude and colours for those stars are given as $\pm$0.02 and $\pm$0.01 for V and B-V, respectively \citep{DH88,F98}. Due to our subfield selection, we had 8-11 standard stars in each field. We then used these standard stars to estimate the standard magnitude of the other stars in the field. 

The magnitude and colour differences between our values and \citet{DH88} and \citet{F98} are shown in Fig. 2. The mean magnitude and colour differences and their respective standard deviations are 0.005, -0.011 and 0.07, 0.07 mag. 

The internal errors, as derived from DAOPHOT, in magnitude and colour are plotted against $V$ magnitude in Fig. 3 and the mean values of the errors are listed in Table 2. Figure 3 shows that photometric error is $\leq0.015$ mag at $V \sim 19$ mag while the colour error is 0.024 mag.   

\begin{table}
\setlength{\tabcolsep}{1pt} 
\center
\caption{Mean photometric errors and number of stars in optical and near-infrared magnitudes and colours for different magnitude intervals. N is the number of the stars for each interval.}
\begin{tabular}{cccccccc}
 &      &      &    & & &  \\
\hline
 V &  $\sigma_{V}$ & $\sigma_{B-V}$ &N$_{BV}$& $\sigma_{J} $ & $\sigma_{J-H}$ & $\sigma_{J-K_{s}}$ & N$_{2MASS}$\\
 (mag) &  (mag) & (mag) & & (mag) & (mag) & (mag) & \\
\hline

(10,11]	& 0.001 & 0.001 &  1& 0.024 & 0.037& 0.033 & 1\\
(11,12]	& 0.001 & 0.001 &  7& 0.023 & 0.036& 0.031 & 7\\
(12,13]	& 0.001 & 0.001 &  9& 0.024 & 0.038& 0.033 & 9\\
(13,14]	& 0.001 & 0.001 & 32& 0.026 & 0.041& 0.036 & 32\\
(14,15]	& 0.001 & 0.001 & 58& 0.026 & 0.041& 0.037 & 58\\
(15,16]	& 0.002 & 0.003 & 70& 0.027 & 0.043& 0.040 & 70\\
(16,17]	& 0.003 & 0.005 &105& 0.032 & 0.051& 0.051 & 105\\
(17,18]	& 0.005 & 0.009 &133& 0.038 & 0.060& 0.067 & 133\\
(18,19]	& 0.009 & 0.016 &194& 0.056 & 0.087& 0.109 & 191\\

\hline
\end{tabular}
\end{table}

\begin{figure}
\centering
   \includegraphics[scale=0.4]{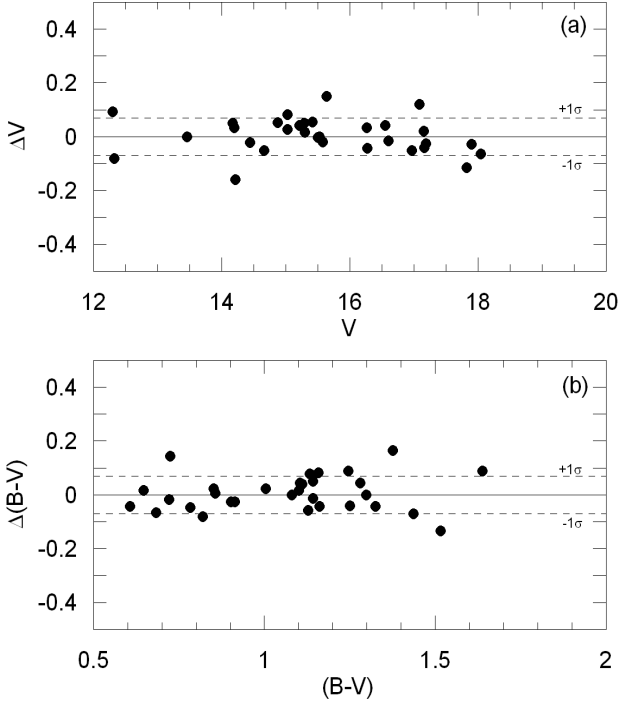}
   \caption{Magnitude (a) and colour (b) differences between the values calculated in this study and Del Rio \& Huestamendia's (1988) and Frandsen \& Arentoft's (1998). $V$ and $(B-V)$ represent the magnitudes and colours obtained in this study. The dashed lines denote $\pm1\sigma$.}
\label{fields}
\end{figure}

\begin{figure}
\centering
   \includegraphics[scale=0.4]{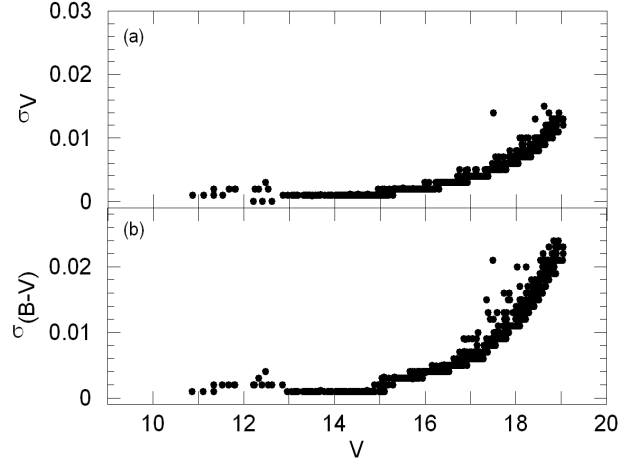}
   \caption{Photometric errors (in magnitude) corresponding to the brightness measurement at $V$ and $(B-V)$ are plotted against the $V$-band brightness. Error on the y-axis represent the internal errors as estimated by DAOPHOT routine.}
\end{figure}

\begin{figure}
\centering
   \includegraphics[scale=0.4]{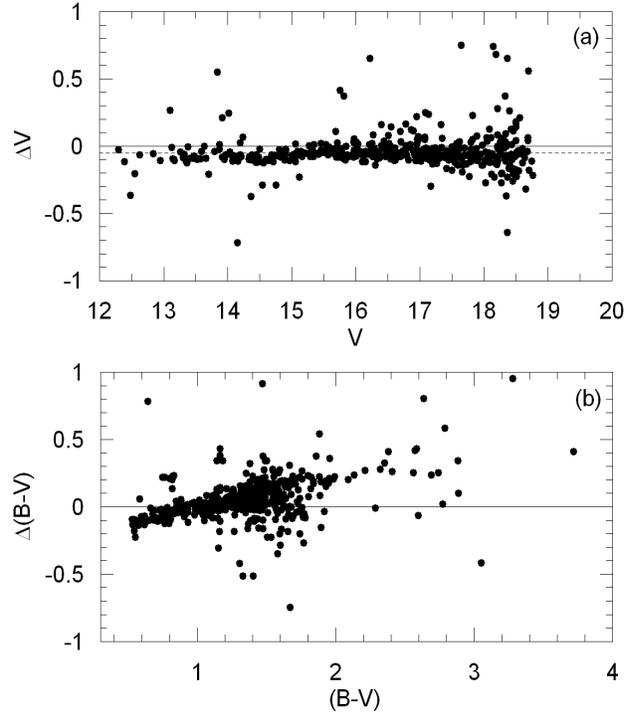}
   \caption{Magnitude (a) and colour (b) differences between the values measured in this study and Maciejewski \& Niedzielski's (2007) values. $V$ and $(B-V)$ represent the magnitudes and colours obtained in this study. The dashed line in panel (a) denotes $\Delta V=-0.05$ mag.}
\label{fields}
\end{figure}

\begin{figure}
\centering
   \includegraphics[scale=0.4]{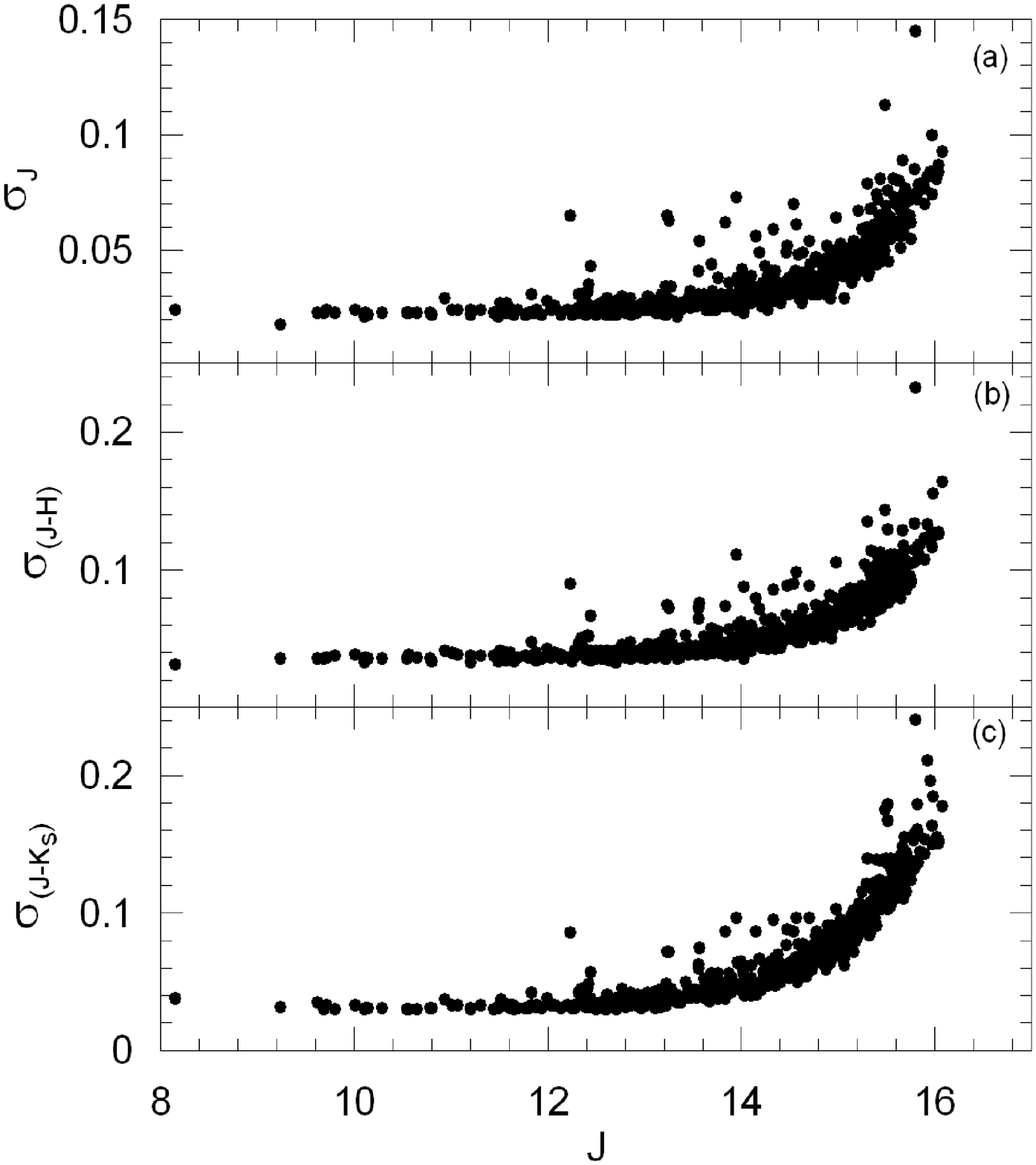}
   \caption{Photometric errors (in magnitude) corresponding to the brightness measurement at $J$, $(J-H)$ and $(J-K_{s})$ are plotted against the $J$-band brightness.}
\end{figure}

We compared our results with a recent BV photometric study on NGC 1513, performed by \citet{MN07}. 452 out of 609 stars in our sample matched with Maciejewski \& Niedzielski's (2007). The magnitude and colour differences between our results and those of Maciejewski \& Niedzielski's (2007) are given in Fig. 4. As seen from Fig. 4a the zero point of the magnitude difference is at $\Delta V=-0.05$ mag. The trend seen in Fig. 4b indicates a disagreement between our values and Maciejewski \& Niedzielski's (2007). After analyzing and comparing our data with previous studies, we conclude that our values are not in agreement with values given by Maciejewski \& Niedzielski (2007). This discrepancy in Fig. 4 might exist because of the fact that Maciejewski \& Niedzielski (2007) did not pick standard stars from a large magnitude and colour interval.

\subsection{Near-Infrared Data}

The near-infrared $JHK_{s}$ photometric data were taken from the digital Two Micron All-sky Survey\footnote{http://www.ipac.caltech.edu/2MASS/} (2MASS). 2MASS uniformly scanned the entire sky in three near-infrared bands $J$ (1.25$\mu$m), $H$ (1.65$\mu$m) and $K_s$ (2.17$\mu$m) with two highly automated 1.3-m telescopes equipped with a three-channel camera, where each channel consists of a $256\times256$ array of HgCdTe detectors. The photometric uncertainty of the data is less than 0.155 at $K_{s} \sim 16.5$ magnitude which is the photometric completeness of 2MASS for stars with $|b| > 25^{o}$ \citep{S06}. We picked stars in a field, of size 25 arcmin$^2$, in the direction of NGC 1513 from Cutri et al.'s Point Source Catalogue (2003) and calculated the limiting magnitudes and photometric errors are $J=16.5 \pm 0.125$, $H=16.0 \pm 0.143$, $K_{s}=15.5 \pm 0.175$ mag. These photometric errors are in agreement with the error values given for high latitude star fields. 

We detected 609 stars in BV photometry in our observations. We then compared our results with Cutri et al.'s Point Source Catalogue (2003) and matched the appropriate stars. After that, we obtained the 2MASS magnitudes for the 606 observed stars. 534 out of these 606 stars have a ``AAA'' quality flag, which means the signal noise ratio is $SNR \geq 10$, i.e. they have the highest quality measurements. The 2MASS magnitude and colour errors are as follows: $0.018\leq\sigma_{J}\leq0.145$, $0.032\leq\sigma_{J-H}\leq0.232$ and $0.030\leq\sigma_{J-K_{s}}\leq0.241$ mag. Errors are given in Table 2 and shown in Fig. 5. Table 2 reveals that the accuracy of optical data is better than near-infrared data, because there are more observations in the former. The coordinates, optical and near-infrared magnitudes and their errors for 609 observed stars are given in Table 3.

\begin{table*}
\setlength{\tabcolsep}{2.1pt} 
\center
\caption{The coordinates, optical and near-infrared magnitudes and their errors for 609 observed stars. The catalogue containing this information can be obtained electronically.}
\begin{tabular}{ccccccccccccc}
 &  &  &  & &  &  &  &  & & & & \\
\hline
ID  & $\alpha_{2000}$    & $\delta_{2000}$    &   B    & B$_{err}$ &    V   & V$_{err}$ & J &   J$_{err}$    & H &   H$_{err}$    & $K_{s}$ &   $K_{{s}_{err}}$ \\
    & (hh:mm:ss)         & (dd:mm:ss)                   &  (mag) & (mag) &    (mag)   & (mag) & (mag) &   (mag)    & (mag) &   (mag)    & (mag) &   (mag) \\

\hline
001 & 04:09:05.16 & 49:30:07.62 & 17.849 & 0.004 & 16.563 & 0.003 & 13.859 & 0.031 & 13.440 & 0.036 & 13.257 & 0.032 \\
002 & 04:09:05.17 & 49:29:11.57 & 20.053 & 0.014 & 18.510 & 0.009 & 15.487 & 0.050 & 14.897 & 0.066 & 14.727 & 0.100 \\
003 & 04:09:05.45 & 49:32:29.04 & 19.473 & 0.010 & 18.108 & 0.007 & 15.384 & 0.055 & 15.065 & 0.071 & 14.627 & 0.086 \\
004 & 04:09:05.72 & 49:25:47.56 & 18.853 & 0.006 & 17.446 & 0.005 & 14.780 & 0.045 & 14.402 & 0.049 & 14.201 & 0.075 \\
005 & 04:09:05.87 & 49:24:24.37 & 19.812 & 0.007 & 17.553 & 0.005 & 15.483 & 0.113 & 15.040 & 0.090 & 14.826 & 0.134 \\
... & ...         & ...         & ...    & ...   & ...    & ...   & ...    & ...   & ...    & ...   & ...    & ...   \\
... & ...         & ...         & ...    & ...   & ...    & ...   & ...    & ...   & ...    & ...   & ...    & ...   \\
... & ...         & ...         & ...    & ...   & ...    & ...   & ...    & ...   & ...    & ...   & ...    & ...   \\
605 & 04:10:29.69 & 49:30:37.04 & 18.652 & 0.005 & 16.921 & 0.003 & 13.892 & 0.029 & 13.158 & 0.036 & 12.952 & 0.030 \\
606 & 04:10:29.72 & 49:27:03.29 & 19.860 & 0.010 & 18.271 & 0.007 & 15.137 & 0.048 & 14.562 & 0.057 & 14.503 & 0.080 \\
607 & 04:10:30.08 & 49:30:28.02 & 16.268 & 0.002 & 15.115 & 0.001 & 13.495 & 0.027 & 13.192 & 0.032 & 13.134 & 0.035 \\
608 & 04:10:31.19 & 49:23:41.85 & 20.263 & 0.004 & 16.545 & 0.003 & 10.005 & 0.024 &  8.608 & 0.031 &  8.195 & 0.023 \\
609 & 04:10:32.86 & 49:26:16.21 & 17.559 & 0.003 & 16.452 & 0.003 & 13.824 & 0.026 & 13.527 & 0.032 & 13.309 & 0.036 \\
\hline
\end{tabular}
\end{table*}

\section{Data Analysis}
\subsection{Cluster's Centre and Radial Density Profile}

NGC 1513 is a cluster with low central concentration. The centre of the cluster can only be determined roughly by eye-estimation. To determine the centre more precisely, we applied the star-count method and divided the cluster into one arcmin sized squares. We then calculated the surface distributions of those squares (Fig. 6). We assumed the centre of the cluster as the square's centre with maximum star density. The star symbol seen in Fig. 6 indicates the centre of the cluster with equatorial coordinates $\alpha_{2000}=04^{h}09^{m}36^{s}$, $\delta_{2000}=49^{\circ}28^{'}43^{''}$ and galactic coordinates $l=152^{\circ}.57$, $b=-1^{\circ}.64$. This result is in agreement with Maciejewski \& Niedzielski's (2007) $\alpha_{2000}=04^{h}09^{m}46^{s}$, $\delta_{2000}=49^{\circ}28^{'}28^{''}$, but somewhat different than Frolov et al.'s (2002) $\alpha_{2000}=04^{h}10^{m}38^{s}$, $\delta_{2000}=49^{\circ}31^{'}00^{''}$ central equatorial coordinates. Considering our results were obtained from both optical and near-infrared photometric systems, we can state that our central coordinates are more accurate.

To establish the radial density profile we counted the stars with distances $r\leq1$ arcmin from the centre of the cluster. We repeated this process up to $r=15$ with 1 arcmin steps. The next step was to subtract the stars in previous areas from the later ones, so that we obtained only the amount of the stars within the relevant area, not a cumulative count. Finally, we divided the star counts in 15 fields by the appropriate areas, i.e. the areas of the fields those stars belong to. The density uncertainties in each field was calculated using Poisson noise statistics. 

Using photometric data we plotted the radial density versus angular separation from the cluster's centre in Fig. 6. The figure shows the radial density profile from the centre of the cluster to a maximum angular separation of 15 arcmin. The density shows a maximum at the centre $\rho=8.5$ stars/arcmin$^{2}$ and then decreases down to $\rho=5.5$ stars/arcmin$^{2}$ at 15 arcmin. The decrease becomes asymptotical at $r\sim10$ arcmin, after that point there are a few cluster stars. To determine the structural parameters of the cluster more precisely, we applied the empirical King model \citep{K66}. The King model parameterizes the density function $\rho(r)$ as:

\begin{equation} 
\rho(r)=f_{bg}+\frac{f_{0}}{1+(r/r_{c})^{2}},
\end{equation} 
where $f_{bg}$, $f_{0}$ and $r_{c}$ are background and central star densities and the core radius of the cluster, respectively. Fig. 6 reveals the background star density $f_{bg}=5.5$ stars/arcmin$^{2}$. We then compared observational values with the ones we obtained from King profile using minimum $\chi^{2}$ statistics. The analysis shows that $f_{0}=3.00\pm0.31$ stars/arcmin$^{2}$ and $r_{c}=3.98\pm0.55$ arcmin. The degree of freedom of the analysis (dof) is 0.11, while its squared correlation coefficient ($R^{2}$) is 0.89. In Fig. 6, the dots stand for the observational density values, whereas solid line represents the King profile. The error bars denote the Poisson errors in observations. 

\citet{MN07} analyzed the structural parameters of 42 open clusters using King model \citep{K66}. They obtained the parameters for NGC 1513 as: $r_{lim}=9.2$, $r_{c}=3.7$ arcmin, $f_{0}=2.47$ and $f_{bg}=1.04$ stars/arcmin$^{2}$. Comparing our results with Maciejewski \& Niedzielski's (2007) we can conclude that although the limiting radii and the core radii are in agreement, the central and background star densities somewhat differ.

To check our background central density value, we used the Besan\c con Galaxy model\footnote{http://model.obs-besancon.fr/} \citep{R03}. We assumed the size of the field as 0.05 deg$^{2}$, and the magnitude range as $10<V<19$. We took colour excess $E(B-V)=0.68$ and distance modulus values $(V-M_{V})=12.95$ mag from Sections 3.3 and 3.4. According to Besan\c con model there are 3.6 field stars per arcmin$^{2}$, which we showed as the dashed horizontal line in Fig. 6. At the core radius $r_{c}=3.98$ arcmin, which we obtained for the cluster using King model \citep{K66}, the background star density is $\rho=5.5$ stars/arcmin$^{2}$. This value differs slightly from Besan\c con model's 3.6 stars/arcmin$^{2}$. The observational and model background star densities are not in total agreement; there is a difference of 1.9 stars/arcmin$^2$. This discrepancy originates from the galactic model parameters, because Galaxy models use several parameters with fixed values for the entire Galaxy. However, \citet{B08} showed that galactic model parameters are a function of absolute magnitude, galactic latitude and longitude, i.e. they do not have fixed values. This means models using fixed galactic model parameters can not always be used to explain observational data in different directions of the Galaxy.

Consequently, the analyses in this section show that the equatorial and galactic coordinates of the centre are $\alpha_{2000}=04^{h}09^{m}36^{s}$, $\delta_{2000}=49^{\circ}28^{'}43^{''}$ and $l=152^{\circ}.57$, $b=-1^{\circ}.64$, respectively, and that the cluster has an angular radius of approximately 10 arcmin.

\begin{figure*}
\centering
   \includegraphics[scale=0.5]{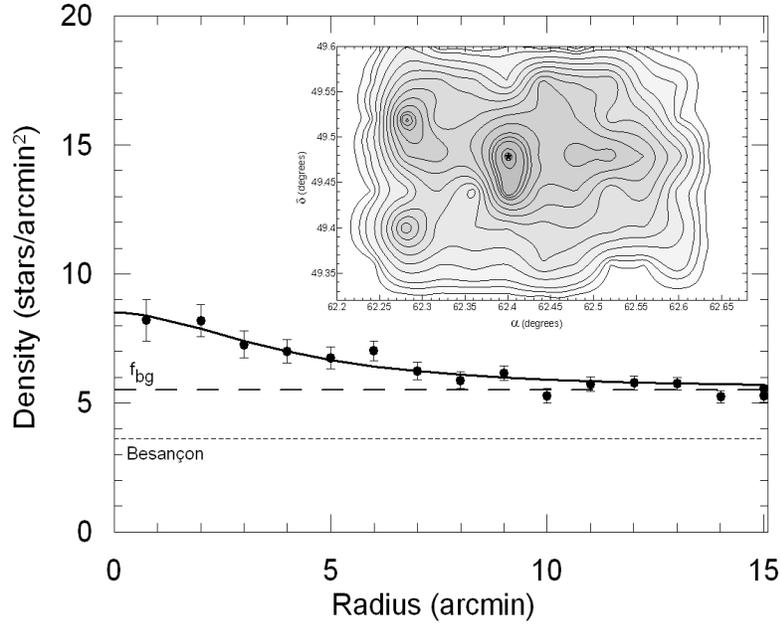}
   \caption{Radial surface distribution of NGC 1513. {\em Dots}: Observational data, {\em solid line}: King profile, {\em dashed line}: Number of background stars calculated using Besan\c con galaxy model, {\em thick dashed line}: Number of observational background stars. Errors are determined from sampling statistics ($1/\sqrt{N}$ where N is the number of stars used in the density estimation at that point). The contour map in the upper right shows how we determined the centre of the cluster. The star symbol in the upper right represents the centre.}
\end{figure*}

\subsection{Colour-Magnitude Diagrams}

We established optical and near-infrared colour magnitude diagrams (CMDs) for NGC 1513. In Fig. 7 we present $(B-V)/V$, $(J-H)/J$ and $(J-K_{s})/K_{s}$ diagrams for NGC 1513. Since the central concentration of the cluster is relatively low, the determination of whether if a star is a field star or a member of the cluster using radial stellar density profile is tough. Even though we calculated the core radius of the cluster as $r_{c}=3.98$ arcmin from Fig. 6, to determine cluster members with even more probability we chose the stars within the circle with radius $r=5$ arcmin. By doing that, we obtained a more precise main-sequence in the CM diagram, which contains 343 stars in our sample. All of these 343 main-sequence stars are found in-between the two dashed lines in Fig. 7a-c. However, there are 110 more stars in-between the dashed lines with $5 < r \leq 15$ arcmin. This means the contamination in Fig. 7 is about 24 \%.

However, Frolov et al.'s (2002) astrometric study regarding 333 stars in the direction of the cluster provides us information about cluster membership of the 609 observed stars. The filled and open circles in Fig. 7 represent Frolov et al.'s (2002) probable cluster members ($p>50\%$) and other observed stars, respectively. The reason we use the probable cluster members is to decide which isochrone to use when determining the age of the cluster. The problem with the astrometrically determined high probability stars is that they can only be detected in brighter magnitudes. As seen in the CM diagrams in Fig. 7, there are a few giants in the direction of the cluster. The cluster membership probability of stars with numbers 360 and 380, which appear in the giant regions of the CMDs, are 90\% and 93\% (Frolov et al., 2002) according to their proper motion studies, respectively. Since there are no spectroscopic studies regarding these systems, it is unknown if these stars are giants or main-sequence stars. Recently, a new method was suggested by \citet{B06} to separate field giants from field dwarfs. This new method is based on the comparison of the 2MASS $J$, $H$, $K_{s}$ with the $V$ magnitudes down to the limiting magnitude of $V=16$ mag. The calibration equations used in separating giants from dwarfs are as follows:

\begin{equation} 
J_{0}=0.957\times V_{0}-1.079,
\end{equation} 

\begin{equation} 
H_{0}=0.931\times V_{0}-1.240,
\end{equation}

\begin{equation} 
(K_{s})_{0}=0.927\times V_{0}-1.292,
\end{equation}
where ``0'' index denotes the de-reddened magnitudes. To apply this method, magnitudes of stars should be de-reddened. The colour excesses and the de-reddening method are given in detail in Section 3.3. The calibration and the de-reddened $V_{0}$, $J_{0}$, $H_{0}$, $(K_{s})_{0}$ magnitudes of the two stars were shown in Fig. 8. The de-reddening procedure is explained in detail in the next section. As seen from the Fig. 8 the stars appear to the right of the calibration lines, which means they are in the giant region. These two stars are cluster members astrometrically and giant stars photometrically. According to Pickles' (1998) synthetic data the giant stars numbered 360 and 380 belong to K1 and K0 spectral types, respectively. 

By selecting the stars within the circle with radius $r=5$ arcmin and using Frolov et al.'s (2002) high probability member stars we determined a more precise main-sequence, which is an important factor in deciding the distance modulus of the cluster. By making this selection, we made a precise determination of the cluster's main-sequence's turn-off point which is the best age indicator. 

\begin{figure*}
\centering
   \includegraphics[scale=0.80]{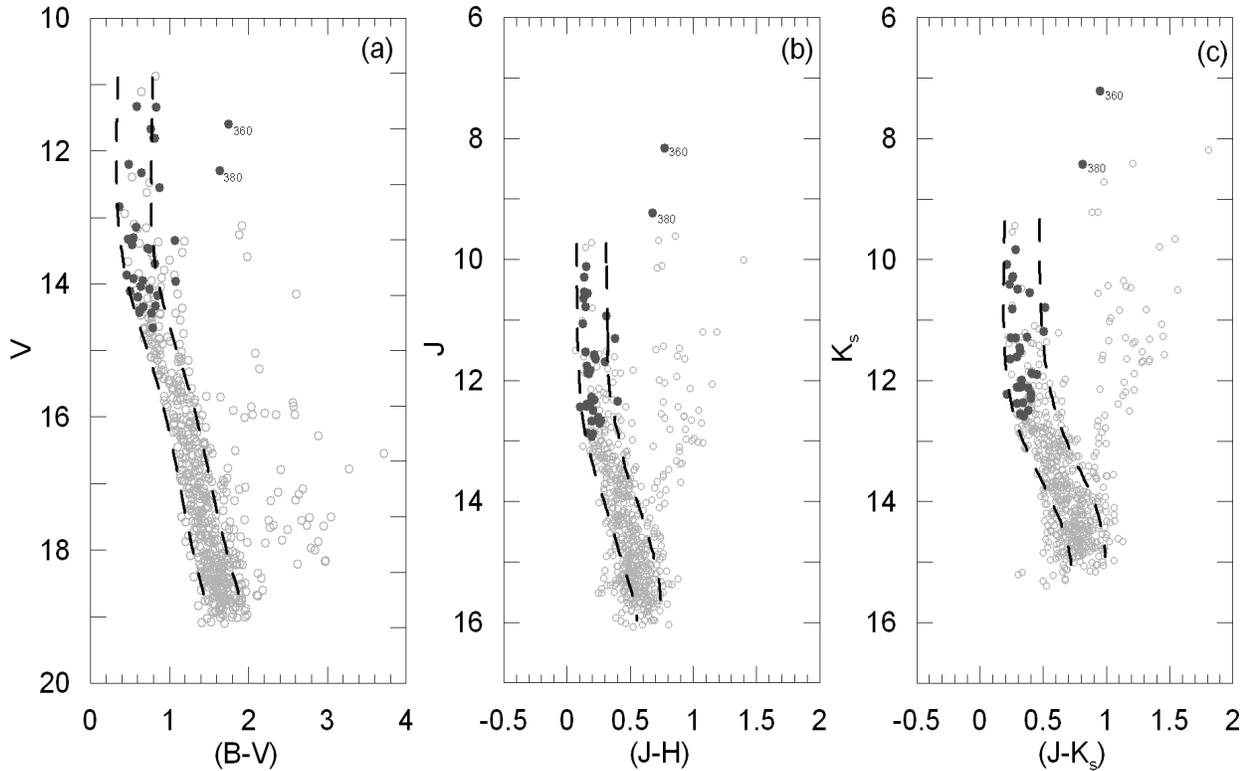}
   \caption{Optical and near-infrared colour-magnitude diagrams for NGC 1513. Empty circles and filled circles denote the stars in this study and Frolov et al.'s (2002) high probability cluster stars, respectively. The dashed lines were obtained using the cluster members with $r \leq 5$ arcmin, while the stars in-between the lines have $r\leq15$ arcmin.} The stars numbered 360 and 380 are also shown in all panels.
\end{figure*}

\begin{figure*}
\centering
   \includegraphics[scale=0.80]{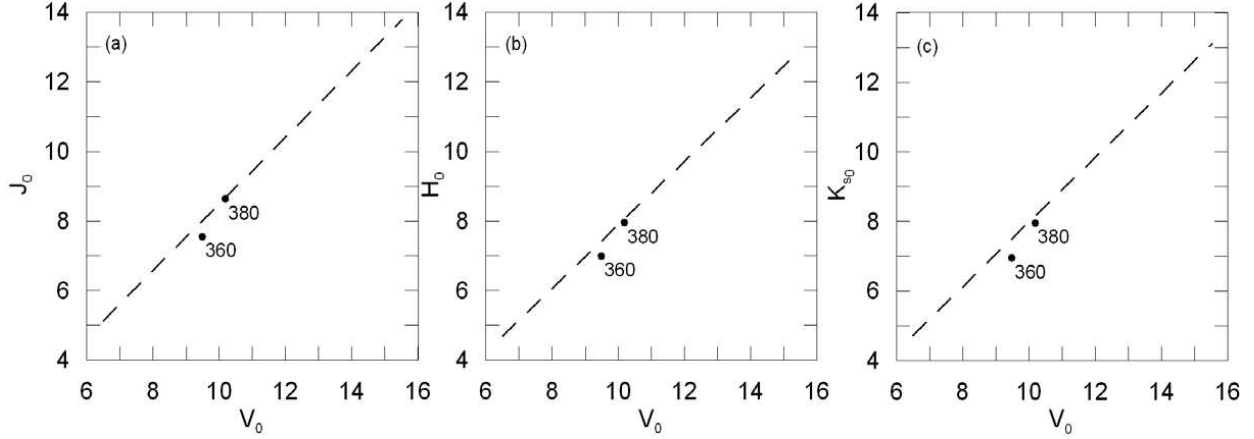}
   \caption{The stars numbered 360 and 380 in magnitude-magnitude diagrams. The dashed lines represent the border of giant/dwarf separation. (a) $V_{0}/J_{0}$, (b) $V_{0}/H_{0}$, (c) $V_{0}/(K_{s})_{0}$.}
\end{figure*}

\begin{table}
\setlength{\tabcolsep}{1.5pt} 
\center
\scriptsize{
\caption{Pickles' (1998) original synthetic colours (1-5). $(J-H)$ and $(J-K_{s})$ were calculated using Carpenter's (2001) transformation equations (6-7).}
\begin{tabular}{ccccccc}
\hline
\multicolumn{5} {c} {$Pickles' colours$} & \multicolumn{2} {c} {$2MASS$}\\
 (1) & (2) &  (3) &  (4) &  (5) & (6) & (7) \\
 SpType & $(B-V)$ &  $(V-J)$ &  $(V-H)$ &  $(V-K)$ & $(J-H)$ & $(J-K_{s})$\\
\hline
       O5V &     -0.380 &     -0.737 &     -0.992 &     -1.096 &     -0.202 &     -0.214 \\
       O9V &     -0.331 &     -0.718 &     -0.899 &     -0.989 &     -0.182 &     -0.194 \\
       B0V &     -0.342 &     -0.695 &     -0.834 &     -0.856 &     -0.163 &     -0.175 \\
       B1V &     -0.244 &     -0.632 &     -0.732 &     -0.780 &     -0.143 &     -0.145 \\
       B3V &     -0.201 &     -0.482 &     -0.734 &     -0.798 &     -0.123 &     -0.115 \\
     B5-7V &     -0.139 &     -0.322 &     -0.370 &     -0.395 &     -0.094 &     -0.076 \\
       B8V &     -0.109 &     -0.229 &     -0.259 &     -0.283 &     -0.074 &     -0.046 \\
       B9V &     -0.044 &     -0.141 &     -0.147 &     -0.160 &     -0.055 &     -0.027 \\
       A0V &      0.015 &      0.003 &      0.000 &     -0.009 &     -0.045 &     -0.017 \\
       A2V &      0.029 &     -0.017 &     -0.149 &     -0.136 &     -0.035 &      0.003 \\
       A3V &      0.089 &      0.036 &     -0.129 &     -0.126 &     -0.016 &      0.032 \\
       A5V &      0.153 &      0.288 &      0.351 &      0.360 &      0.014 &      0.062 \\
       A7V &      0.202 &      0.369 &      0.459 &      0.474 &      0.043 &      0.101 \\
       F0V &      0.303 &      0.533 &      0.663 &      0.683 &      0.082 &      0.140 \\
       F2V &      0.395 &      0.607 &      0.621 &      0.645 &      0.122 &      0.190 \\
       F5V &      0.458 &      0.827 &      1.030 &      1.058 &      0.180 &      0.248 \\
       F6V &      0.469 &      0.900 &      1.110 &      1.128 &      0.210 &      0.288 \\
       F8V &      0.542 &      1.012 &      1.219 &      1.245 &      0.249 &      0.317 \\
       G0V &      0.571 &      1.017 &      1.275 &      1.300 &      0.298 &      0.376 \\
       G5V &      0.686 &      1.190 &      1.452 &      1.507 &      0.288 &      0.386 \\
       K2V &      0.924 &      1.650 &      2.142 &      2.226 &      0.445 &      0.563 \\
       K3V &      0.930 &      1.808 &      2.366 &      2.396 &      0.484 &      0.612 \\
       K4V &      1.085 &      1.973 &      2.541 &      2.643 &      0.523 &      0.661 \\
       K5V &      1.205 &      2.172 &      2.770 &      2.872 &      0.553 &      0.691 \\
       K7V &      1.368 &      2.398 &      2.968 &      3.094 &      0.602 &      0.779 \\
       M0V &      1.321 &      2.855 &      3.514 &      3.678 &      0.612 &      0.809 \\
       M1V &      1.375 &      2.974 &      3.622 &      3.896 &      0.602 &      0.909 \\
       M2V &      1.436 &      3.294 &      3.961 &      4.139 &      0.602 &      0.829 \\
       M3V &      1.515 &      3.817 &      4.449 &      4.674 &      0.582 &      0.839 \\
       M4V &      1.594 &      4.422 &      5.038 &      5.306 &      0.563 &      0.860 \\
       M5V &      1.663 &      5.253 &      6.096 &      6.393 &      0.563 &      0.920 \\
       M6V &      1.816 &      6.362 &      7.026 &      7.409 &      0.602 &      1.008 \\
\hline

\end{tabular}
}  
\end{table}

\subsection{Two Colour Diagrams and Colour Excesses}

We present optical and near-infrared two-colour diagrams for $r\leq5$ arcmin in Fig. 9. In Fig. 9a, we plotted $(J-H)_{0}$ versus $(B-V)_{0}$ whereas in Fig. 9b the axes are $(J-K_{s})_{0}$ and $(B-V)_{0}$. Namely, we plotted a near-infrared colour versus an optical colour in each panel. To determine the reddening in the direction of the cluster we made use of the synthetical library of \citet{P98}. In Pickles' (1998) library we selected the metallicity to be $[M/H]=0$ dex and main-sequence stars of different spectral types and obtained datasets for $(B-V)$, $(V-J)$, $(V-H)$, $(V-K)$. Since Pickles' near-infrared data are in Johnson's system, we used Carpenter's (2001) transformation equations (A1, A2, A3, A4) to obtain magnitudes of 2MASS bands (Table 4). We plotted the standard main-sequence from Table 4 and our observations in two-colour diagrams (Fig. 9). In order to determine the reddening we calculated loci (represented by star symbols in Fig. 9) for our observational data and plot the best fit for those loci. These loci represent the main-sequence for our observations and we slide that main-sequence in both directions until it fits best with the standard main-sequence. The amount of sliding gives us the colour excesses for Fig. 9a and b, $E(J-H)$, $E(B-V)$ and $E(J-K_{s})$, $E(B-V)$ respectively. The colour excesses and their relative errors we obtained using minimum $\chi^{2}$ method are as follows: $E(J-H)=0.21\pm0.02$, $E(B-V)=0.68\pm0.06$ and $E(J-K_{s})=0.33\pm0.04$, $E(B-V)=0.68\pm0.06$ mag, respectively. The confidence level of colour excess errors is 99.5\%. The contour maps of two-colour diagrams in Fig. 10 show the optimum colour excesses for BV and 2MASS photometries.

\citet{DH88} obtained RGU colour excess as $E(G-R)=0.94$ mag from standard stars. \citet{F02} converted this value into $E(B-V)$ using Steinlin's (1968) formula and calculated $E(B-V)=0.67$ mag. The recent study of \citet{MN07} estimates $E(B-V)=0.76^{+0.13}_{-0.18}$ mag. Obviously, our value is almost in perfect agreement with Frolov et al.'s (2002), whereas Maciejewski \& Niedzielski's (2007) agrees only within error bars.

\citet{FM03} calculated the colour excess values for 2MASS photometric system. According to them the colour excess ratios are: $E(J-H)/E(B-V)=0.322\pm0.074$ and $E(J-K_{s})/E(B-V)=0.505\pm0.053$ mag. We ended up with the following results: $E(J-H)/E(B-V)=0.309\pm0.130$ and $E(J-K_{s})/E(B-V)=0.485\pm0.150$ mag. The results produced for normal interstellar medium by \citet{FM03} is in agreement with our results.

\begin{figure*}
\centering
   \includegraphics[scale=0.7]{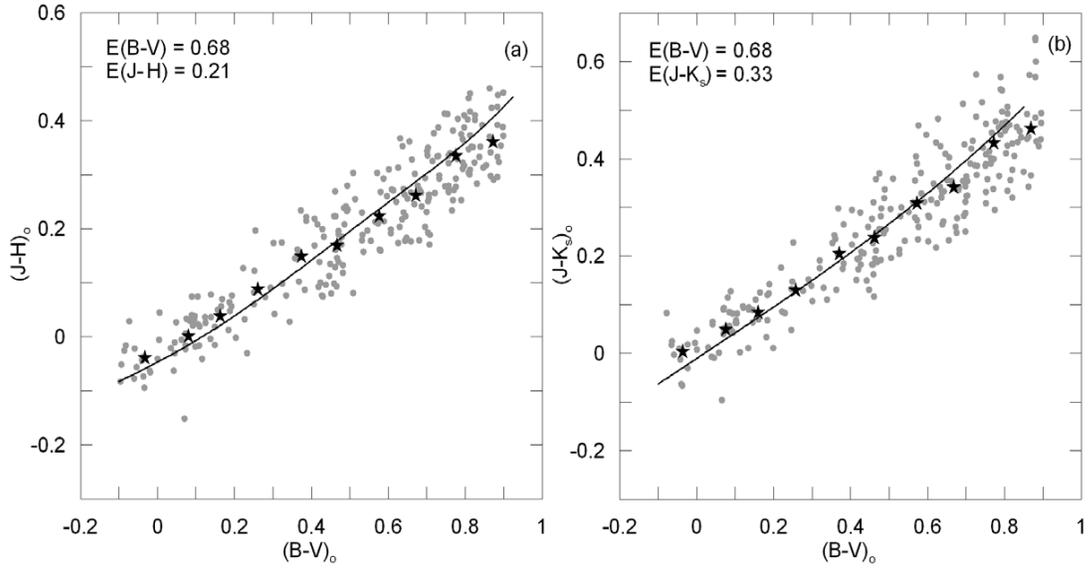}
   \caption{Optical and near-infrared two-colour diagrams. Filled circles, the solid line and star symbols represent stars within $r\leq5$ arcmin angular separation from the cluster centre, Pickles' (1998) original synthetic main-sequences and the loci of the main-sequence stars for the cluster, respectively.}
\end{figure*}

\begin{figure*}
\centering
   \includegraphics[angle=0, width=150mm, height=74mm]{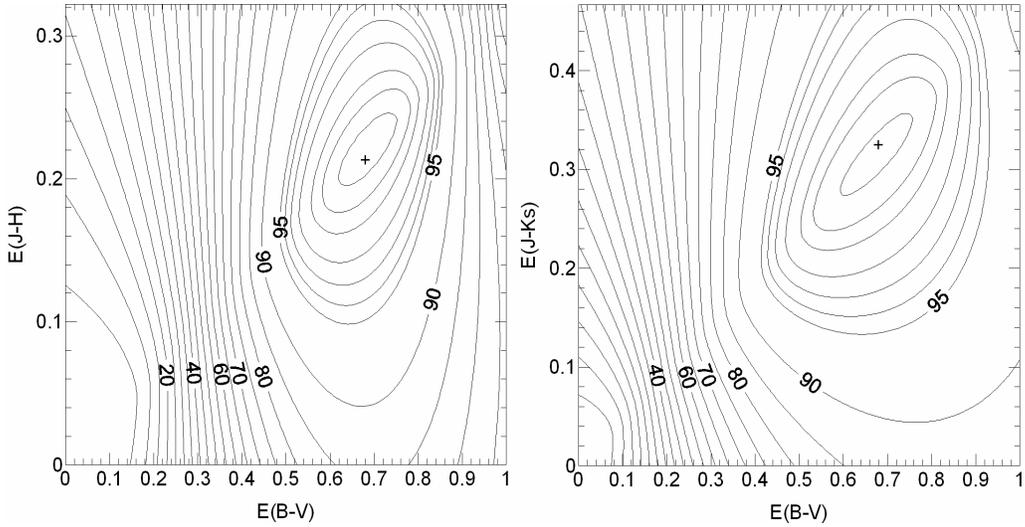}
   \caption{Colour excess diagrams. The plus symbol represents the reduced minimum $\chi^{2}$ in both panels, whereas the contours denote the probability distribution of reduced $\chi^{2}$. These contours are used in determining the colour excesses and their relative errors.}
\end{figure*}

\begin{figure*}
\centering
   \includegraphics[scale=0.80]{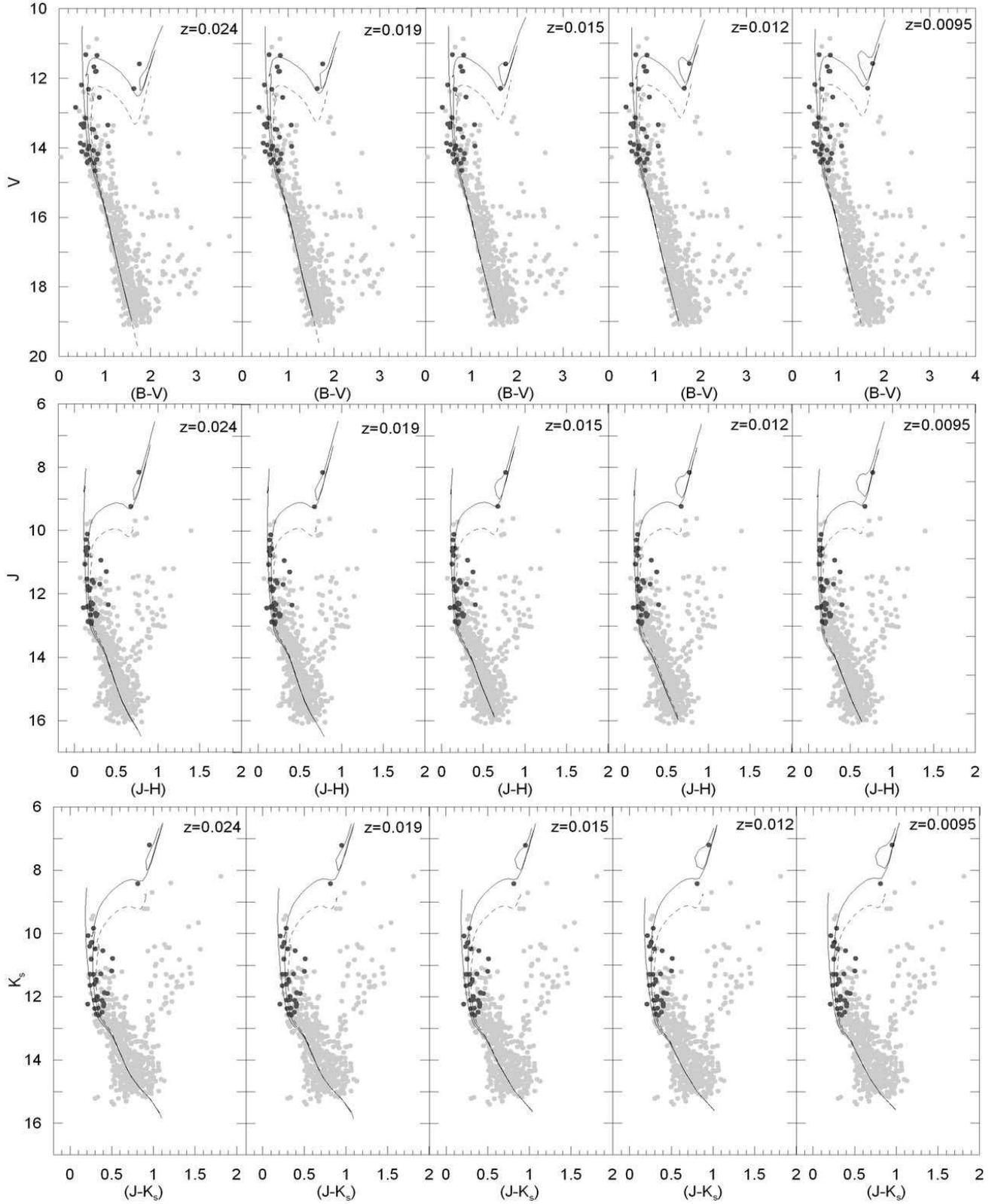}
   \caption{$V/(B-V)$, $J/(J-H)$ and $J/(J-K_{s})$ CMDs of the cluster. The black circles represent high probability ($P > 50\%$) cluster members; whereas the grey circles denote the low probability cluster members and field stars. Thin solid line: ZAMS, thick solid curve line: $\log (t/yr)=8.40$, thin dashed curve line: $\log (t/yr)=8.60$.}
\end{figure*}

\begin{figure*}
\centering
   \includegraphics[scale=0.80]{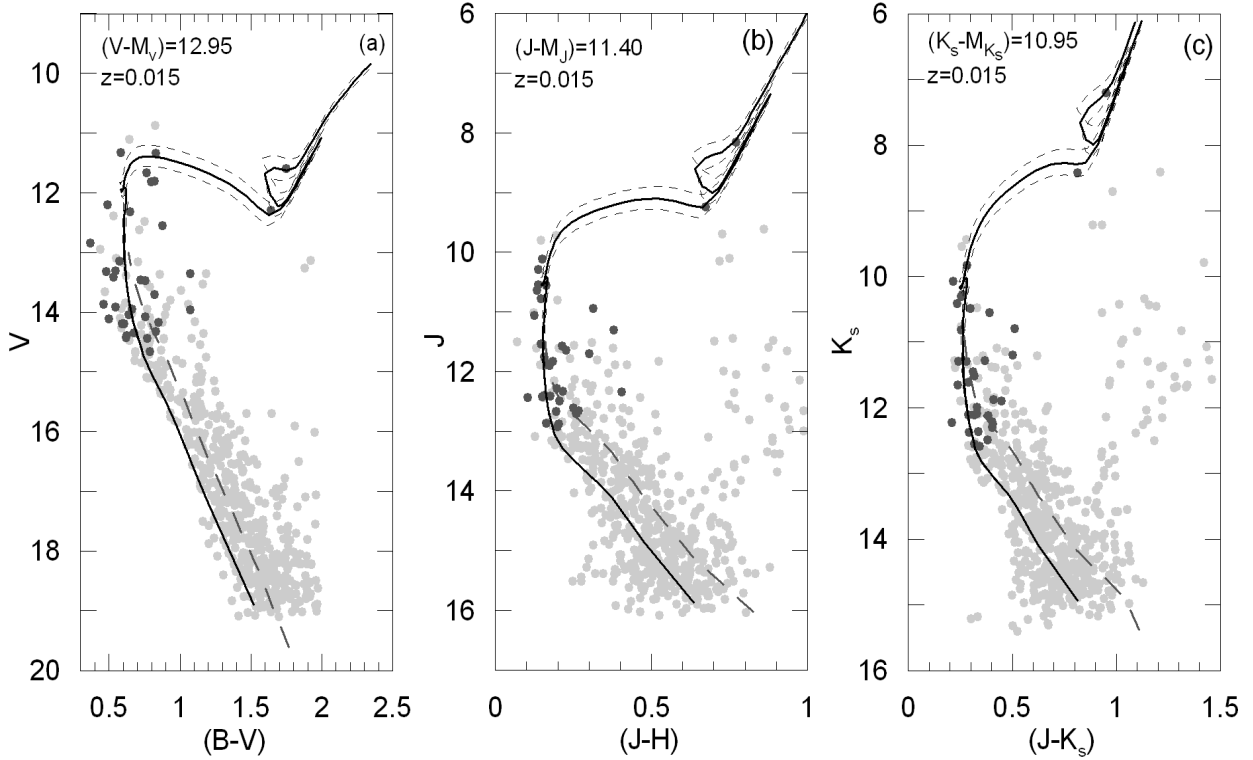}
   \caption{$V/(B-V)$, $J/(J-H)$ and $J/(J-K_{s})$ CMDs of the cluster. The solid, dashed upper and lower curves correspond to $\log (t/yr)=8.40$, $\log (t/yr)=8.35$ and $\log (t/yr)=8.44$ isochrones, respectively. Finally, the thick-dashed line stands for binary sequence.}
\end{figure*}

\subsection{Distance and Age of the Cluster}

The zero age main-sequence (ZAMS) fitting procedure was used to derive the distance to the cluster. We added Schmidt-Kaler's (1982) ZAMS to the optical CMDs in Fig. 11. We used the colour excess $E(B-V)=0.68$ mag discussed in Section 3.3 and reddened Schmidt-Kaler's (1982) ZAMS accordingly. We slid the cluster's main-sequence vertically until it overlapped with the \citet{SK82} one. The distance modulus of the cluster is the amount of sliding we have applied to the cluster's main-sequence, which is $(V-M_{V})=12.95$ mag. Since there is no 2MASS data regarding Schmidt-Kaler's (1982) ZAMS, we have selected proper 2MASS data from the Padova isochrones\footnote{http://stev.oapd.inaf.it/cgi-bin/cmd} with solar metallicity. We plotted 2MASS ZAMS \citep{M08} in Fig. 11. Then, we applied the previous procedure to 2MASS CMDs (Fig. 11) using reddening values of $E(J-H)=0.21$ and $E(J-K_{s})=0.33$ mag. The optical and near-infrared ZAMS are given as thin solid lines in all panels of Fig. 11. Finally, we calculated the distance moduli for $J/(J-H)$ and $K_{s}/(J-K_{s})$ diagrams and obtained $(J-M_{J})=11.40$ and $(K_{s}-M_{K_{s}})=10.95$ mag, respectively. To de-redden the distance moduli calculated from CMDs, we used Fiorucci \& Munari's (2003) formulae: $A_{V}=3.1\times E(B-V)$, $A_{J}=0.887\times E(B-V)$, $A_{K_{s}}=0.322\times E(B-V)$ and obtained: $(V-M_{V})_{0}=10.84\pm0.19$, $(J-M_{J})_{0}=10.80\pm0.10$ and $(K_{s}-M_{K_{s}})_{0}=10.73\pm0.10$ mag, using these de-reddened distance moduli we obtained the distance of NGC 1513 1472, 1445 and 1400 pc, respectively. These results, obtained from two different photometric systems, are of $\pm 5\%$ relative difference with each other. Moreover, both \citet{F02} and \citet{MN07} calculated the distance of the cluster as 1320 pc, which indicates a relative difference of $\simeq 10\%$.

The age of a star cluster can be determined by comparing the observed CMDs with theoretical isochrones. To determine the age of a cluster colour excess, distance modulus and metallicity needs to be known. In this study, we determined the colour excess and distance modulus, separately. However, since there is no spectroscopic data regarding the NGC 1513 its metallicity remains unknown. Hence, we took into account the metallicity during the calculation of the age by using a set of isochrones for stars with masses $0.15<M_{\odot}\leq100$, different metal abundance $0.0001\leq Z\leq 0.03$ and ages from $\log (t/yr)<10.24$ published on the web site of the Padova research group and described in the work of \citet{M08}. We produced several isochrones for optical and near-infrared bands with different metal abundances ranging from 0.0095 to 0.024, which corresponds to $[M/H]=-0.30$ and $[M/H]=0.10$ dex, respectively (solar abundance was assumed as $Z_{\odot}=0.019$). The CMDs and the isochrones are given in Fig. 11. On each CMD, we plotted three isochrones representing three different ages: ZAMS, $\log (t/yr)=8.40$, $\log (t/yr)=8.60$. As seen from Fig. 11, the isochrones with $Z=0.015$ provide the best fit for the cluster's main-sequence, main-sequence turn-off point and giant stars. Therefore, we assumed $Z=0.015$ (corresponds to a metallicity of $[M/H]=-0.10$ dex) to be the metal abundance of the cluster. 

To determine the age precisely, we plotted three isochrones with ages $\log (t/yr)=8.35$, $\log (t/yr)=8.40$ and $\log (t/yr)=8.44$ in the CMDs with $Z=0.015$ (Fig. 12). The isochrone with $\log (t/yr)=8.40$ seemed to represent the cluster best, because it fits with both the main-sequence and the giants of the cluster. The loop (Fig. 12a: $B-V=0.61$, $V=11.93$; Fig. 12b: $J-H=0.16$, $J=10.44$; Fig. 12c: $J-K_{s}=0.26$, $K_{s}=10.05$ mag) in the $\log (t/yr)=8.40$ isochrone marks the ending of hydrogen burning in the core, compression of the core and hydrogen burning in the thick layer. The star numbered 380 is located almost at the base of the red giant branch, and the star numbered 360 – at the stage of helium burning. The turn-off point of the main-sequence corresponding to this isochrone is $(B-V)_{0}=-0.11$ mag, which is the colour index for the B8 spectral type. To determine the binary effect regarding the cluster, we assumed the stars in the cluster to be binary systems of equal-massed components. In this case, absolute magnitude is 0.75 mag brighter than it should normally be. This binary main-sequence with $\log (t/yr)=8.40$ has been plotted in Fig. 12 as a thick-dashed line.

Consequently, the average distance of the cluster is $d=1440\pm80$ pc, the metallicity $Z=0.015\pm0.004$ ($[M/H]=-0.10\pm 0.10$ dex) and average age is $\log (t/yr)\sim8.40$. These results partially agree with Frolov et al.'s (2002) results. They determined the distance, metallicity and average age as $d=1320$ pc, $Z=0.019$ ($[M/H]=0$ dex) and $\log (t/yr)=8.40$. Comparing our results with Frolov et al.'s (2002), we can claim that the distance we calculated is within 10\% of each other, while the metallicity seems lower and age is the same. \citet{MN07} estimated the age of the cluster $\log (t/yr)=7.4$, which is not in agreement with the age we calculated, which might be due to the fact that they did not take the giant stars of the cluster into account. To determine the integrated absolute magnitudes and colours of NGC 1513 in optical and near-infrared, we used Lata et al.'s (2002) equations and obtained the following results for cluster stars ($r\leq5$ arcmin): $I(M_{V})=-2.136$, $I(B-V)_{0}=0.032$, $I(J-H)_{0}=0.053$ and $I(J-K_{s})_{0}=-0.037$ mag. According to Table 4 the optical integrated colour corresponds to A2 spectral type.

\section{Conclusion}

We present CCD BV and JHK$_{s}$ 2MASS photometric data for the low central concentration young star cluster NGC 1513. The results obtained in the analysis are the following:

i) We determined the centre of the cluster as $\alpha_{2000}=04^{h}09^{m}36^{s}$, $\delta_{2000}=49^{\circ}28^{'}43^{''}$ and its galactic coordinates $l=152^{\circ}.57$, $b=-1^{\circ}.64$. The radial density profile shows that the angular radius of the cluster is $r=10$ arcmin.

ii) The optical and near-infrared colours of the cluster main-sequence reveal the colour excesses, $E(B-V)=0.68\pm0.06$, $E(J-H)=0.21\pm0.02$ and $E(J-K_{s})=0.33\pm0.04$ mag. We estimated $E(J-H)/E(B-V)=0.309\pm0.130$ and $E(J-K_{s})/E(B-V)=0.485\pm0.150$ mag.

iii) We compared the optical main sequence of the cluster with Schmidt-Kaler's (1982) ZAMS, the near-infrared one with Padova isochrones' ZAMS \citep{M08}. We obtained the distance moduli as $(V-M_{V})_{0}=10.84\pm0.19$ for optical colour and $(J-M_{J})_{0}=10.80\pm0.10$, $(K_{s}-M_{K_{s}})_{0}=10.73\pm0.10$ mag for near-infrared colours. The average distance derived from these moduli is $1440\pm80$ pc. 

iv) The Padova isochrone with $Z=0.015$ and $\log (t/yr)=8.40$ provides the best fit for the cluster in both optical and near-infrared CMDs. Therefore, we conclude that the cluster is $\log (t/yr)=8.40\pm0.04$ and has a metallicity of $[M/H]=-0.10\pm0.10$ dex.

\section{Acknowledgements}
We thank to T\"UB\.ITAK for a partial support in using RTT150 (Russian-Turkish 1.5-m telescope in Antalya) with project number TUG-RTT150.04.016. The anonymous referee's contributions towards the paper helped us improve it. We would also like to thank Dr. Funda G\"uver and Astronomer Hikmet \c Cakmak for their contributions. This research has made use of the WEBDA database, operated at the
Institute for Astronomy of the University of Vienna. This research has made use of the SIMBAD database, operated at CDS, Strasbourg, France. 
This publication makes use of data products from the 2MASS, which is a joint project of the University of Massachusetts and the Infrared 
Processing and Analysis Center/California Institute of Technology, funded by the National Aeronautics and Space Administration and the 
National Science Foundation.

\end{document}